\documentclass[aip,apl,reprint,a4paper,floatfix,amsmath,amssymb,amsfonts,noshowpacs,citeautoscript]{revtex4-2}
\usepackage{newtxtext,newtxmath}
\usepackage[T1]{fontenc}
\usepackage{graphicx}
\usepackage[dvipsnames]{xcolor}
\usepackage{soul} 
\usepackage{floatrow}
\usepackage[
,textwidth=17.5cm
,textheight=23.5cm
,verbose
,pdftex
]{geometry}
\usepackage{xspace}

\usepackage[pdftex]{hyperref}
\hypersetup{
  pdftitle={AlScNmultilayer},
	colorlinks,
	citecolor=blue,
	linkcolor=blue
	}

\makeatletter
\newcommand*{\refig}[2]{\hyperref[#1]{\ref*{#1}(#2)}}
\makeatother

\DeclareMathAlphabet{\mathsf}{OT1}{\sfdefault}{m}{n}
\SetMathAlphabet{\mathsf}{bold}{OT1}{\sfdefault}{b}{n}
\DeclareUnicodeCharacter{2212}{-}

\begin{document}
\graphicspath{{./figures/}}
\title{Lattice-Matched Multiple Channel AlScN/GaN Heterostructures}

\author{Thai-Son~Nguyen}
\email[Electronic mail: ]{tn354@cornell.edu}
\affiliation{\hbox{Department of Materials Science and Engineering, Cornell University, Ithaca, New York 14853, USA}}
\author{Naomi~Pieczulewski}
\affiliation{\hbox{Department of Materials Science and Engineering, Cornell University, Ithaca, New York 14853, USA}}
\author{Chandrashekhar~Savant}
\affiliation{\hbox{Department of Materials Science and Engineering, Cornell University, Ithaca, New York 14853, USA}}
\author{Joshua~J. P. Cooper}
\affiliation{\hbox{Department of Materials Science and Engineering, University of Michigan, Ann Arbor, Michigan 48109, USA}}
\author{Joseph~Casamento}
\affiliation{Department of Materials Science and Engineering, Massachusetts Institute of Technology, \\Cambridge, Massachusetts 02139, USA}
\author{Rachel~S. 
Goldman}
\affiliation{\hbox{Department of Materials Science and Engineering, University of Michigan, Ann Arbor, Michigan 48109, USA}}
\author{David A.~Muller}
\affiliation{\hbox{Kavli Institute at Cornell for Nanoscale Science, Cornell University, Ithaca, New York 14853, USA}}
\affiliation{\hbox{School of Applied and Engineering Physics, Cornell University, Ithaca, New York 14853, USA}}
\author{Huili G.~Xing}
\affiliation{\hbox{Department of Materials Science and Engineering, Cornell University, Ithaca, New York 14853, USA}}
\affiliation{\hbox{Kavli Institute at Cornell for Nanoscale Science, Cornell University, Ithaca, New York 14853, USA}}
\affiliation{\hbox{Department of Electrical and Computer Engineering, Cornell University, Ithaca, New York 14853, USA}}
\author{Debdeep~Jena}
\affiliation{\hbox{Department of Materials Science and Engineering, Cornell University, Ithaca, New York 14853, USA}}
\affiliation{\hbox{Kavli Institute at Cornell for Nanoscale Science, Cornell University, Ithaca, New York 14853, USA}}
\affiliation{\hbox{School of Applied and Engineering Physics, Cornell University, Ithaca, New York 14853, USA}}
\affiliation{\hbox{Department of Electrical and Computer Engineering, Cornell University, Ithaca, New York 14853, USA}}

\begin{abstract}
AlScN is a new wide bandgap, high-k, ferroelectric material for RF, memory, and power applications. Successful integration of high quality AlScN with GaN in epitaxial layer stacks depends strongly on the ability to control lattice parameters and surface or interface through growth. This study investigates the molecular beam epitaxy growth and transport properties of AlScN/GaN multilayer heterostructures. Single layer Al$_{1-x}$Sc$_x$N/GaN heterostructures exhibited lattice-matched composition within $x$ = 0.09 -- 0.11 using substrate (thermocouple) growth temperatures between 330 $ ^\circ$C and 630 $ ^\circ$C. By targeting the lattice-matched Sc composition, pseudomorphic AlScN/GaN multilayer structures with ten and twenty periods were achieved, exhibiting excellent structural and interface properties as confirmed by X-ray diffraction (XRD) and scanning transmission electron microscopy (STEM). These multilayer heterostructures exhibited substantial polarization-induced net mobile charge densities of up to 8.24 $\times$ 10$^{14}$/cm$^2$ for twenty channels. The sheet density scales with the number of AlScN/GaN periods. By identifying lattice-matched growth condition and using it to generate multiple conductive channels, this work enhances our understanding of the AlScN/GaN material platform.
\end{abstract}

\maketitle

\section{\label{sec:level1}Introduction}

Aluminum scandium nitride (AlScN) is an emerging group III-nitride material that has received significant research interest in the past decade. The addition of Sc into AlN was first reported to drastically increase the piezoelectric coefficients of AlScN alloy compared to the highly piezoelectric AlN.\cite{Akiyama_2009_Influencegrowthtemp_ScAlNpiezo, Akiyama_2009_EnhancementPiezoelectric_AlScN_sputtered} More recently, reactive sputtered and plasma-assisted molecular beam epitaxy (PA-MBE) AlScN films have been reported to be the first ferroelectric III-nitride material.\cite{Fichtner_2019_ferroelectric_alscn, Wang_2022_EpitaxialFerroelectric_ScAlN-GaN_memory, zhang_nanoscale_2023, yasuoka_tunable_2022, uehara_lower_2022, kim_impact_2023} AlScN also shows high potential in nonlinear optics\cite{Yoshioka_2021_opticalnonlinearity_AlScN} and distributed Bragg reflectors\cite{van_deurzen_2023_AlScN_DBR}. These properties of AlScN add new functionalities by device integration with both III-nitride semiconductors, and with Silicon CMOS where it is already in use in commercial BAW filters. 

While the growth, structural, electrical, and piezoelectric properties of single-layer AlScN thin films have been extensively studied, the integration of AlScN multilayers remains unexplored. Multilayers of binary, ternary, and quaternary alloys in group III-Arsenides, and in group III-Nitrides have enabled various electronic and photonic device applications such as multichannel high-electron mobility transistors (HEMTs), resonant tunneling diodes, solar cells, laser diodes, and infrared emitters and detectors. In the III-As material family, lattice-matched (Al,Ga)As alloys allowed for the development of single- and multi-layer heterostructures such as multiple quantum wells and superlattices with low dislocation density and excellent interfaces\cite{kroemer_nobel_2001} which enabled device technologies such as superlattice-castellated field-effect transistors (SLCFETs)\cite{Howell_2023_multichannel_GaN_SLCFET_10.87W-mm_94GHz} and quantum cascade lasers.\cite{faist_quantum_1994}

Lattice-matched group III-Nitride heterostructures are limited due to the 2.6\% in-plane lattice mismatch of AlN--GaN and 13\% lattice mismatch of AlN-InN. Al$_{0.82}$In$_{0.18}$N/GaN is the only known lattice-matched ternary-binary pair and can be difficult to grow due to the large growth temperature mismatch between InN and AlN. Using Al(Ga,In)N/GaN multilayer structures, GaN-based resonant tunneling diodes\cite{kikuchi_alngan_2002}, distributed Bragg reflectors\cite{butteRecentProgressGrowth2005}, quantum cascade lasers\cite{hirayama_recent_2015}, multichannel HEMTs for high-power electronic and RF switches\cite{Xiao_2020_3.3kV_multichannel_AlGaN-GaN_SBD,Howell_2023_multichannel_GaN_SLCFET_10.87W-mm_94GHz} have been realized. However, structural degradation due to strain relaxation\cite{Terano_2015_gan-based_multi2DEG_3kV} and complex strain engineering of Al(Ga)N/GaN superlattices\cite{waldripStressEngineeringMetalorganic2001,huangCrackfreeGaNAlN2006} pose intrinsic limitations in such multilayers.

AlScN is a promising candidate to help overcome these limitations because (1) lattice-matching of AlScN to GaN is theoretically predicted and experimentally demonstrated to be possible,\cite{Moram_2014_ScGaNandScAlN_emergingnitride, Dinh_2023_latticematch_AlScN_MBE,Deng_2013_bandgap_AlScN, van_deurzen_2023_AlScN_DBR} (2) AlScN has a large growth temperature window that is compatible with GaN epitaxy\cite{Hardy_2020_phasepurity_highSc_ScAlN_MBE, Casamento_2020_structural-piezo_ultrathinScAlN, Wang_2020_MBE_characterization_ScAlN,motoki_improved_2024}, making it compatible with both CMOS- and GaN-based growth requirements, and (3) AlScN offers strong spontaneous polarization\cite{Akiyama_2009_EnhancementPiezoelectric_AlScN_sputtered, Ambacher_2023_review_ScAlN}, high refractive index mismatch to GaN\cite{van_deurzen_2023_AlScN_DBR, Wang_2020_MBE_characterization_ScAlN,maeda_structural_2024}, and compressive or tensile strain compositions for flexible bandgap, refractive index, and strain engineering. 

The growth and characterization of AlScN/GaN multilayer heterostructures have been studied by a few groups recently. Dzuba et al examined epitaxial AlScN/GaN multiple quantum well heterostructures with quantum well thicknesses below 10 nm to study near-infrared intersubband absorption\cite{Dzuba_2022_Elimination_phaseimpurity_AlScN}. Lattice-matched AlScN/GaN multilayer heterostructures were demonstrated as a highly reflective distributed Bragg reflector\cite{van_deurzen_2023_AlScN_DBR}. However, the lattice-matched composition of AlScN on GaN has been reported to range from 9\% to 20\% Sc\cite{Hardy_2020_phasepurity_highSc_ScAlN_MBE, Casamento_2020_structural-piezo_ultrathinScAlN, Dinh_2023_latticematch_AlScN_MBE, Dzuba_2022_Elimination_phaseimpurity_AlScN, Hoglund_2010_wurtziteScAlN_reactivesputtering, Dargis_2020_single-crystal_multilayer_nitride,motoki_improved_2024,kumar2024pinpointinglatticematchedconditionswurtzite}. Finding the correct lattice-matched composition of AlScN on GaN and its dependence on growth methods and conditions is critical to realize high-quality AlScN/GaN multilayer structures. 

This work aims to (1) identify and use the correct Sc composition for lattice-matched AlScN on GaN to (2) realize lattice-matched AlScN/GaN multilayer heterostructures, and (3) study polarization-induced charges in these multilayers. To achieve these goals, (1) we grew AlScN thin films of 80-100 nm thickness on bulk GaN substrates. By varying the Sc composition and growth temperature, we find the lattice-matched composition of MBE AlScN on GaN to be between 9\% and 11\% Sc composition. (2) Using the single layer lattice-matched condition, we demonstrate pseudomorphic AlScN/GaN multilayer heterostructures on semi-insulating GaN substrates that exhibit high crystalline and interface quality as we verify by X-ray diffraction (XRD) and scanning transmission electron microscopy (STEM). We measure an in-plane lattice mismatch of 0.02\% between AlScN and GaN epilayers in a ten-period 12\% Sc AlScN/GaN sample. (3) These pseudomorphic multilayer AlScN/GaN heterostructures were observed to house extremely high mobile carrier densities exceeding 8 $\times$ 10$^{14}$/cm$^2$ for twenty 2DEG channels. The sheet carrier density was seen to scale linearly with the number of repeating AlScN/GaN periods. 

\section{\label{sec:level2}Methods}

We grew AlScN single-layer and multilayer heterostructures in a Veeco GenXplor MBE reactor. Scandium (99.9\% purity, Ames Laboratory), aluminium (99.9999\% purity), gallium (99.99999\% purity) and silicon (99.99999 purity) were supplied using Knudsen effusion cells. We used an RF plasma source with a nitrogen flow rate of 1.95 sccm and 200~W RF power to provide active nirogen species. We monitored thin film growth in situ using a KSA Instruments reflection high energy electron diffraction (RHEED) apparatus with a Staib electron gun operating at 15 kV and 1.5 A. The growth temperatures were measured by a thermocouple on the backside of the substrate. We used bulk silicon-doped Ga-polar n-type GaN (n$^{+}$GaN) from Ammono for single layer AlScN growths, and bulk semi-insulating GaN substrate from Ammono for multilayer AlScN/GaN growth to evaluate the transport properties of mobile charges in this heterostructure. All substrates were c-plane and Ga-polar, yielding c-oriented, metal-polar MBE GaN and AlScN films.

We grew AlScN films under nitrogen-rich conditions with a metal (Sc+Al) to nitrogen (III/V) ratio of 0.7 to achieve wurtzite phase AlScN\cite{Hardy_2017_epitaxial_ScAlN,Casamento_2020_structural-piezo_ultrathinScAlN, Hardy_2020_phasepurity_highSc_ScAlN_MBE, Wang_2020_MBE_characterization_ScAlN}. We employed metal-rich conditions for GaN epitaxy to promote step-flow growth mode. The beam equivalent pressures (BEPs) of Sc, Al, and Ga used were $\sim$1.5 $\times$ 10$^{-8}$ torr, $\sim$1 $\times$ 10$^{-7}$ torr, and $\sim$6 $\times$ 10$^{-7}$ torr, respectively. The growth rates of AlScN and GaN were 3.0 nm/min and 3.8 nm/min, respectively.  

To explore the effect of Sc composition on the structural properties of AlScN films, we grew 80-100 nm thick AlScN films on 100~nm Si-doped n$^{+}$GaN buffer layer on bulk n$^{+}$GaN substrates. Since the critical thickness of AlScN on GaN is affected by Sc composition and strain relaxation, only films within $\pm$ 1.3\% Sc composition from the lattice-matched composition can remain coherently strained for AlScN thicknesses between 80 and 100 nm\cite{Moram_2014_ScGaNandScAlN_emergingnitride,zhang_2013_Elastic_constants_critical_thickness_ScAlN}, limiting false identification of strained films as the lattice-matched compositions. Unless otherwise noted, we grew single layer AlScN films at 530~$^{\circ}$C and n$^{+}$GaN buffer layers at 630~$^{\circ}$C. 

We grew one ten-period and one twenty-period 45 nm AlScN/40 nm GaN multilayer structure on bulk semi-insulating GaN to evaluate the structural qualities of multilayer AlScN/GaN heterostructures with thick AlScN/GaN periods, using the same III/V ratios of $\sim$ 1 and 0.7 for GaN and AlScN, respectively, and a substrate thermocouple temperature of 530~$^{\circ}$C. Similar heterostructures with the same AlScN and GaN thicknesses have demonstrated high reflectivity as distributed Bragg reflectors.\cite{van_deurzen_2023_AlScN_DBR} All layers were grown at 530~$^{\circ}$C substrate thermocouple temperature. We fully desorbed excess Ga in the GaN epilayers before growing subsequent AlScN layers to prevent Ga incorporation during the nitrogen-rich AlScN growth.

A PANalytical Empyrean system with Cu K$_{\alpha1}$ radiation was used for X-ray diffraction (XRD) and reciprocal space mapping (RSM). We characterized the surface morphology using an Asylum Research Cypher ES atomic force microscope (AFM). To determine the scandium composition, we employed energy-dispersive X-ray spectroscopy using a Zeiss LEO 1550 FESEM equipped with a Bruker energy dispersive X-ray spectroscopy (EDS) silicon drift detector (SDD), X-ray photoelectron spectroscopy (XPS) using aScienta-Omicron-ESCA-2SR XPS instrument equipped with a 1486.6 eV Al K$\alpha$, and Rutherford backscattering spectrometry (RBS) at the Michigan Ion Beam Laboratory using 2 MeV $\alpha$ particles generated in a General Ionex tandem accelerator with backscattered $\alpha$ particles detected by a silicon surface-barrier detector located at 170° with respect to the incident beam. We used a standard van der Pauw geometry and a Nanometrics Hall HL5500 system with a magnetic field of 0.325 T to measure the transport properties of AlScN/GaN multilayers at room temperature. We prepared cross-sectional STEM samples using a Thermo Fisher Helios G4 UX Focused Ion Beam with a final milling step of 5 keV and carried out STEM measurements using an aberration-corrected Thermo Fisher Spectra 300 CFEG operated at 300 keV. Secondary ion mass spectrometry (SIMS) measurement was performed by Eurofins Evans Analytical Group (EAG).

\section{\label{sec:level3}Results and Discussion}


\begin{figure}[t]
\includegraphics[width=8cm]{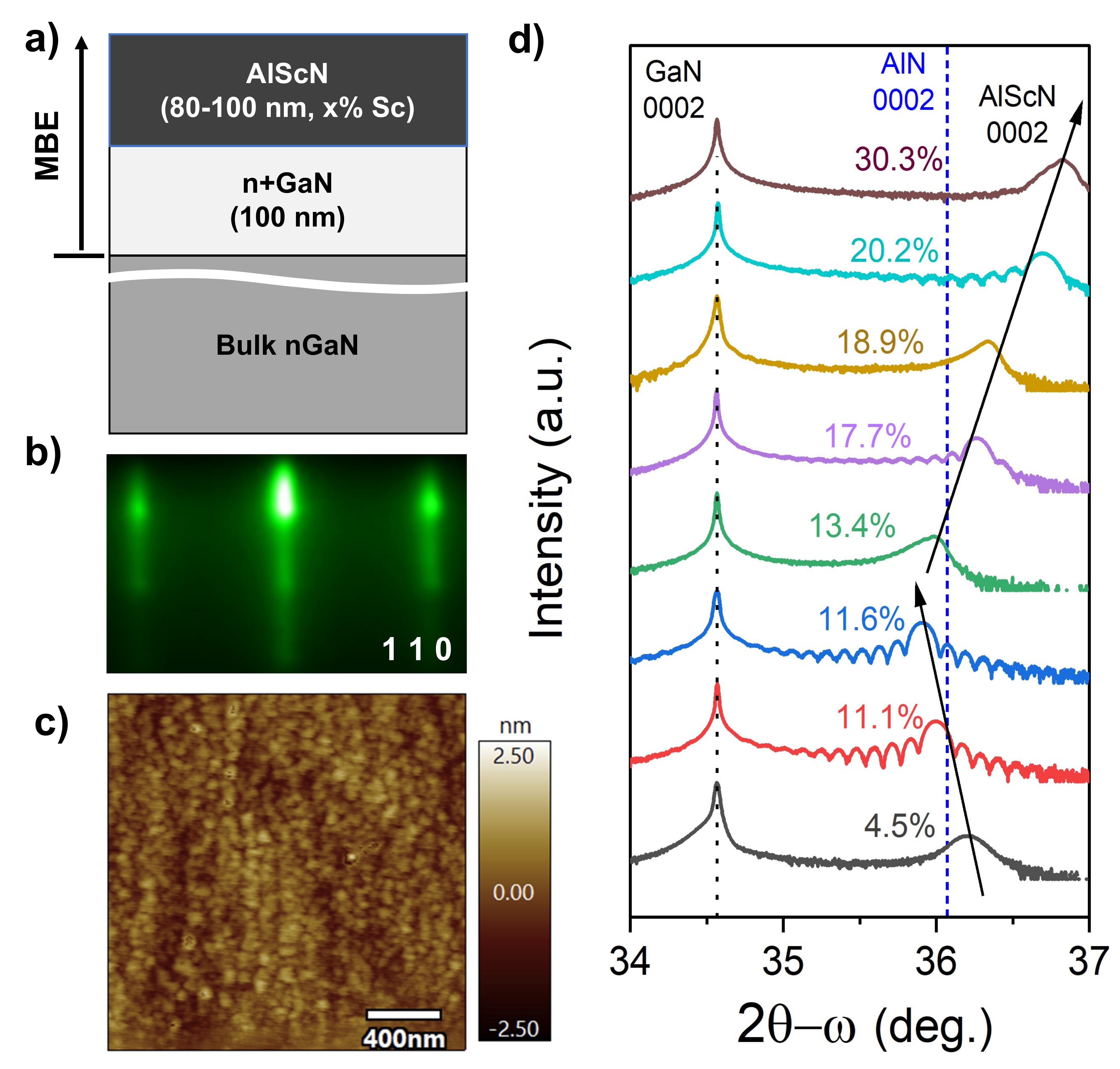}
\caption{(a) Single layer AlScN films with thickness around 100 nm and varying Sc compositions were grown on bulk n-type GaN substrate. (b) Typical RHEED pattern of an AlScN layer grown under nitrogen-rich conditions showing spot-modulated streak patterns. (c) A typical 2$\times$2 $ \mu$m$^2$ AFM micrograph of a 100-nm AlScN layer with rms roughness below 1 nm. (d) Symmetric 2$\theta$-$\omega$ XRD scans depict the non-monotononic change in the c-parameter of AlScN with Sc composition. \label{fig:Figure 1}}
\end{figure}


Figure~\ref{fig:Figure 1}(a) shows the layer structure of Al$_{1-x}$Sc$_x$N films of thickness 80 - 100 nm grown on bulk n$^{+}$GaN substrates to determine the Sc content $x$ that satisfies the lattice-matching condition to GaN, and to explore the crystal quality and surface morphology as $x$ is varied. We chose this thickness range to achieve relaxed films on the bulk nGaN substrate. Figure~\ref{fig:Figure 1}(b) shows a typical in situ RHEED pattern along the <110> zone axis after AlScN growth. The spot-modulated streak pattern suggests a mixed growth mode in which island-like surface features are promoted due to the N-rich growth condition. Figure~\ref{fig:Figure 1}(c) shows a typical 2$\times$2 $ \mu$m$^2$  post-growth surface morphology measured by AFM, with reasonably smooth root mean square (rms) roughness of 0.5 to 1 nm. The surface is dominated by 30-50 nm diameter island features, consistent with the RHEED pattern and other reports on N-rich growth of AlScN \cite{Hardy_2017_epitaxial_ScAlN, Casamento_2020_structural-piezo_ultrathinScAlN}. Figure~\ref{fig:Figure 1}(d) shows symmetric 2$\theta$-$\omega$ scans that confirm the wurtzite phase with strong c-axis orientation of AlScN films via the strong (0002) AlScN diffraction peak near 2$\theta$ = 36$^{\circ}$. The shift from high to low to high 2$\theta$ peak values depicts the non-monotonic dependence of the out-of-plane $c$ lattice parameter with Sc composition. 

The non-monotonic change of the AlScN peak with Sc content $x$ indicated by arrows was reported in previous studies\cite{Casamento_2020_structural-piezo_ultrathinScAlN, Hardy_2020_phasepurity_highSc_ScAlN_MBE}. It is attributed to the distorted bond angles and lengths in the (Al,Sc)N$_4$ tetrahedra\cite{Urban_2021_first-principles_electroacoustic_ScAlN_distortion}. This non-monotonic trend also suggests there are \textit{two} alloy compositions between 9\% and 15\% Sc that match the out-of-plane $c$ lattice parameter of AlN, indicating the potential for future integration of non-polar epitaxial AlScN in m-plane and a-plane AlN-based photonic devices.\cite{bryan_homoepitaxial_2014} The crystalline wurtzite phase can be maintained to $\sim$ $x$ = 0.3, or 30\% Sc with interference fringes observed in Figure \ref{fig:Figure 1}(d) for some compositions, indicating that highly smooth AlScN-GaN interface can be achieved over a broad range of Sc compositions. Notably, the AlScN (0002) peak with more than five satellite Pendell\"{o}sung fringes on each side and strongest relative fringe intensity were observed in samples with $x$ $\approx$ 0.11 -- 0.12, suggesting the highest interface quality between AlScN and n$^{+}$GaN layers achieved near these compositions. 


\begin{figure}[t]
\includegraphics[width=7cm]{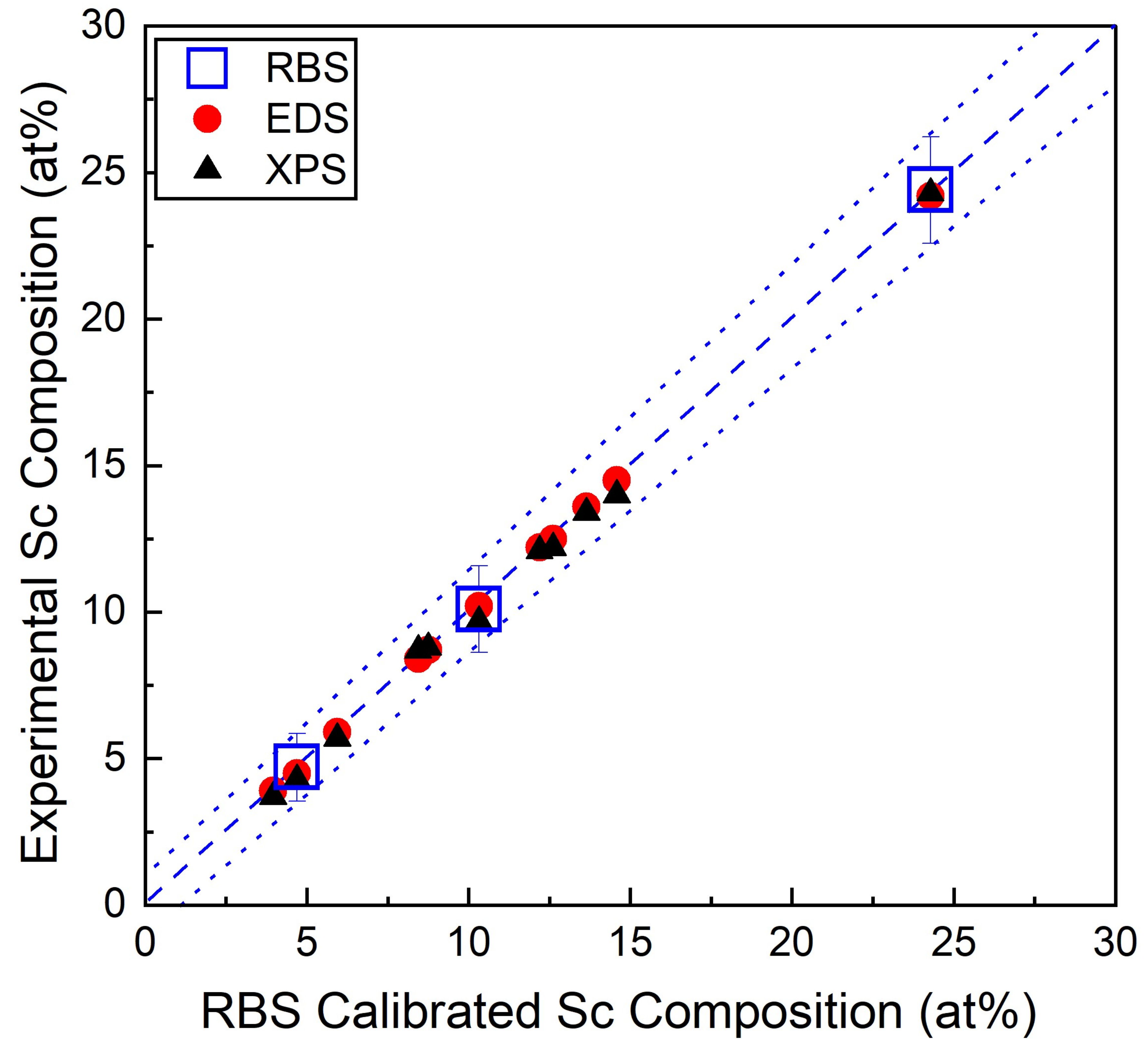}
\caption{Sc content in single layer AlScN films measured by Rutherford backscattering spectrometry (RBS), Energy dispersive X-ray spectroscopy (EDS), and X-ray photoelectron spectroscopy (XPS). Three samples were measured by RBS, a standardless method, and served as calibration standards for EDS and XPS. The RBS measurement uncertainty is shown for reference.The Sc compositions measured in eight additional samples by XPS and EDS closely follow the RBS linear-fit (dashed line). The dotted lines show the $\pm$ 1.5\% range of uncertainty based on RBS measurements.} 
\label{fig:Figure 2}
\end{figure}


\begin{figure*}[t]
\includegraphics[width=\textwidth]{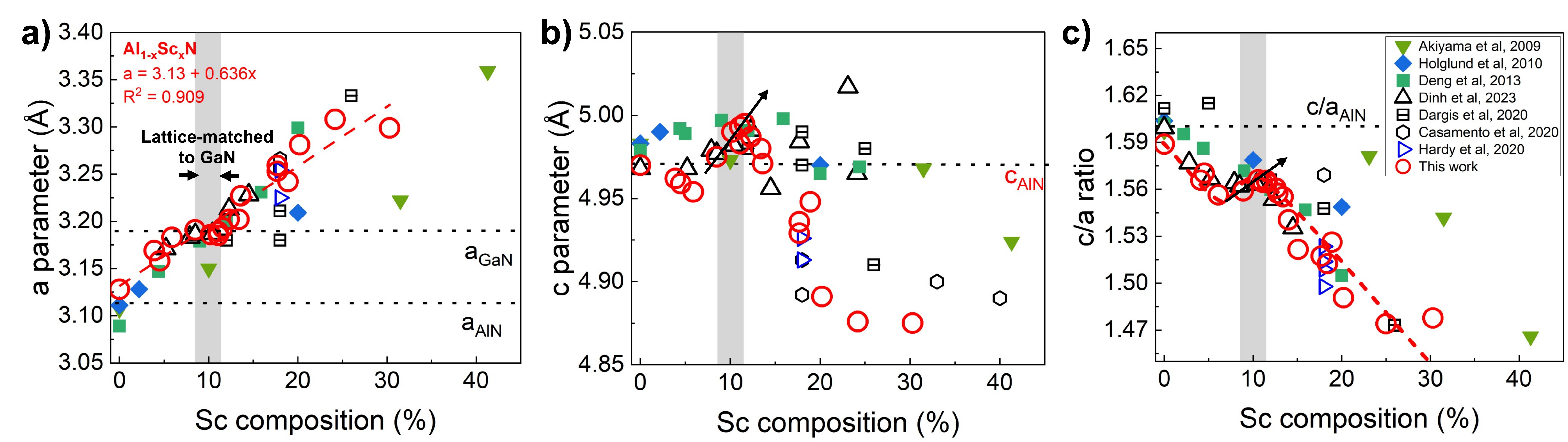}
\caption{Lattice parameters of Al$_{1-x}$Sc$_x$N samples grown at 530 $^{\circ}$C vs. Sc composition obtained from RSM scans (hollow red circles), compared to values reported in previous studies with AlScN grown by MBE (hollow)\cite{Dinh_2023_latticematch_AlScN_MBE, Dargis_2020_single-crystal_multilayer_nitride, Casamento_2020_structural-piezo_ultrathinScAlN, Hardy_2020_phasepurity_highSc_ScAlN_MBE} and reactive sputtering (solid)\cite{Akiyama_2009_Influencegrowthtemp_ScAlNpiezo, Hoglund_2010_wurtziteScAlN_reactivesputtering, Deng_2013_bandgap_AlScN}. (a) AlScN in-plane lattice parameter $a$ increases monotonically with increasing Sc composition, showing clear strain relaxation outside of 9\% - 11\% Sc. The linear fit equation was performed on the data from this work (red circles). (b) AlScN out-of-plane lattice parameter c vs. Sc composition. The grey stripe highlights how the c-parameter increases with Sc content around the lattice-matched composition. (c) $c/a$ ratio shows a monotonically decreasing trend except between 9\% and 11\% Sc. A dashed line (a) and arrows (b, c) are added as guides to the eye. The results in this study align most closely with the data from Dinh et al\cite{Dinh_2023_latticematch_AlScN_MBE}. 
\label{fig:Figure 3}}
\end{figure*}

To accurately determine the Sc content in the AlScN films, we used the complementary experimental techniques of RBS, EDS, XPS, and SIMS. Of these techniques, RBS, EDS, and SIMS provide thin film composition information, while XPS provides surface composition information. We first measured three samples over a broad range of Sc composition (5--25\%) by RBS, EDS, and XPS, with RBS compositions serving as the standard. Sc compositions were determined from RBS yield versus energy spectra, with $\alpha$ backscattered from Sc detected at energies between 1.3 and 1.45 MeV. Analysis of the RBS data was conducted using the simulation of nuclear reaction analysis (SIMNRA) software. For XPS, a hemispherical capacitor analyzer was used to collect the photoelectrons to obtain high-resolution Al 2p, Sc 3s, Sc2p, and N 1s spectra. For EDS, the Al K$\alpha$ and Sc K$\alpha$ peaks were used to determine the Sc composition. It is important to use a standardless method like RBS to calibrate EDS, sensitive to accelerating voltage conditions, and XPS, sensitive to relative sensitivity factors and surface impurity.

Figure~\ref{fig:Figure 2} shows EDS and XPS Sc compositions plotted against the RBS compositions for three samples with Sc composition of 4.7 $\pm$ 1.2\% Sc, 10.1 $\pm$ 1.5\%, and 24.4 $\pm$ 1.8\% Sc measured by RBS. The same samples measured by EDS at 12 kV (XPS) showed Sc compositions of 4.5 $\pm$ 0.2\% (4.3 $\pm$ 0.4\%), 10.3 $\pm$ 0.1\% (10.3 $\pm$ 0.6\%), and 24.3 $\pm$ 0.3\% (24.5 $\pm$ 0.3\%) Sc composition, demonstrating good agreement with the RBS composition. The EDS and XPS Sc compositions of eight additional AlScN samples without RBS data also showed a good correlation between EDS and XPS compositions and closely tracked the RBS linear-fit (dashed line), suggesting the reliability of both methods after calibration. Because EDS is less sensitive to surface contamination than XPS and less time-consuming than RBS and SIMS, it is the method of choice in most Al$_{1-x}$Sc$_x$N studies to date with the uncertainty of $x$ $\sim$ $\pm$ 0.01, or 1 \%.\cite{Dinh_2023_latticematch_AlScN_MBE, Fichtner_2019_ferroelectric_alscn, Wang_2020_MBE_characterization_ScAlN} Unless otherwise noted, the Sc compositions reported in this work were determined by EDS with an expected uncertainty of $\pm$ 1.5\% Sc mainly due to the RBS uncertainty. 

Next, we examined the in-plane lattice constant $a$, out-of-plane lattice constant $c$, and the $c/a$ ratio of Al$_{1-x}$Sc$_x$N samples grown at 530 $^{\circ}$C using RSM as plotted in Figure \ref{fig:Figure 3}. Figure~\ref{fig:Figure 3}(a) shows that $a$ increases monotonically with increasing Sc composition. As highlighted by the grey box in Figure ~\ref{fig:Figure 3}(a), $a_{AlScN}$ $\approx$ $a_{GaN}$ = 3.189 $\Angstrom$ for $x$ $\sim$ 0.09 -- 0.11, and strain relaxation occurs outside this range as suggested by RSM (supplementary, Figure~S1). These near lattice-matched compositions correspond well with the samples that exhibited strong thickness fringes in symmetric 2$\theta$-$\omega$ scans in Figure~\ref{fig:Figure 1}(d). Linear fitting of the $a$ parameter data in this work with Sc composition yielded $a$ = 3.13 + 0.636$x$, where $x$ is the Sc mole fraction in Al$_{1-x}$Sc$_x$N, suggesting a lattice-matched composition of 9.4 $\pm$ 1.5\% Sc. Figure~\ref{fig:Figure 3}(b) depicts the AlScN out-of-plane lattice constant $c$ vs. Sc content. For 0 $\leq x \leq$ 0.08 and 0.13 $\leq x \leq$ 0.30, $c$ decreases with increasing Sc composition. However, $c$ increases quickly for 0.08 $< x <$ 0.13, in accordance with the non-monotonic trend observed in the symmetric 2$\theta$-$\omega$ scans [Figure~\ref{fig:Figure 1}(d)]. Notably, this increase of $c$ corresponds directly to the compositions around which $a_{AlScN}$ $\approx$ $a_{GaN}$, i.e. the lattice-matched composition. 

While the non-monotononic change in $c$ with Sc composition is mainly attributed to AlScN tetrahedron distortion, the strong correlation between the increase of $c$ and the near-lattice-matched compositions observed here suggests a contribution from the Poisson's ratio of the thin film near the lattice-matched composition. With increasing Sc content, a pseudomorphic AlScN film becomes more compressively strained in the in-plane direction and tensilely strained in the out-of-plane direction, therefore causing the $c$ parameter to increase with Sc\%. The $c/a$ ratio in Figure~\ref{fig:Figure 3}(c) also shows a monotonically decreasing trend, except between $x$ = 0.09 and $x$ = 0.11, i.e. the near lattice-matched compositions.

\begin{figure*}[t]
\includegraphics[width=\textwidth]{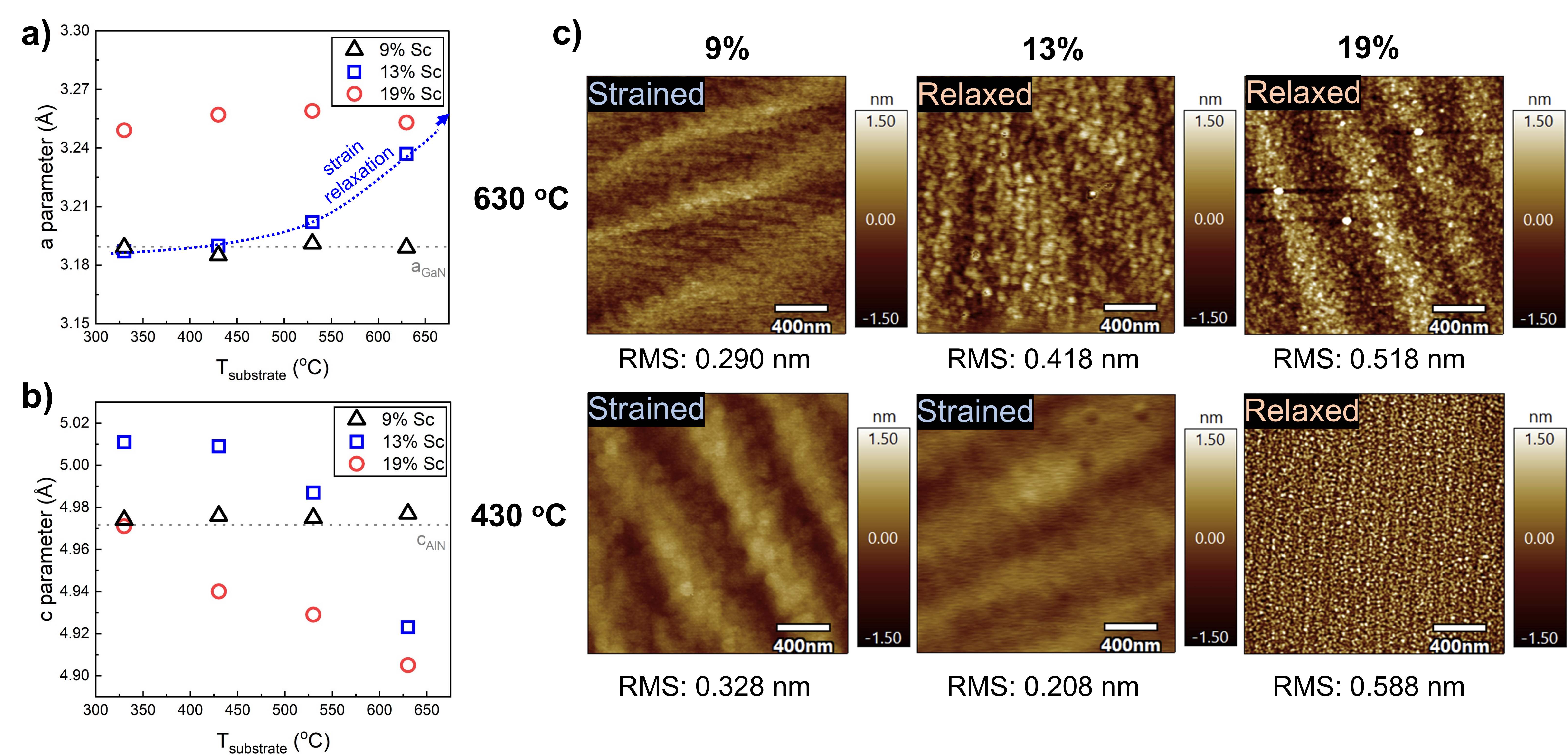}
\caption{(a) AlScN in-plane a-parameter and (b) out-of-plane c-parameter as a function of growth temperature. Similar Sc composition and III/V ratio were used. The a-parameter of 13\% Sc films increases quickly with temperature due to thermally-induced strain relaxation at higher growth temperatures. (c) 2$\times$2 $ \mu$m$^2$ AFM micrographs of samples grown at 630 $ ^\circ$C and 430 $ ^\circ$C thermocouple temperature. Step-flow growth mode can be achieved for near lattice-matched (9\% Sc) samples grown at higher temperatures. As Sc composition and growth temperature increase, the morphology becomes more 3D-dominant, and the rms roughness increases.\label{fig:Figure 4}}
\end{figure*}


It is important to note that previous studies have reported several lattice-matched Sc compositions ranging from 9\% to 20\%~Sc.\cite{Hardy_2020_phasepurity_highSc_ScAlN_MBE, Casamento_2020_structural-piezo_ultrathinScAlN, Dinh_2023_latticematch_AlScN_MBE, Dzuba_2022_Elimination_phaseimpurity_AlScN, Hoglund_2010_wurtziteScAlN_reactivesputtering, Dargis_2020_single-crystal_multilayer_nitride} We see in Figure~\ref{fig:Figure 3} that MBE AlScN films have a more drastic increase in $a$, decrease in $c$, and decrease in $c/a$ ratio compared to sputtered AlScN films. This observation can have strong implications on the difference in wurtzite-rocksalt transition boundary, structural and piezoelectric properties, and ferroelectric properties of MBE and sputtered AlScN\cite{Hardy_2020_phasepurity_highSc_ScAlN_MBE, Yasuoka_2021_ImpactDepositionTemperature_CrystalandFerro_AlScN, Akiyama_2009_Influencegrowthtemp_ScAlNpiezo,Fichtner_2019_ferroelectric_alscn}. AlScN film thickness can affect the measured in-plane lattice parameter due to strain relaxation. The lattice-matched composition between 9 and 11\% Sc determined from Figure~\ref{fig:Figure 3} is in good agreement with Dinh et al's report of 9\% Sc lattice-matched composition of MBE-grown AlScN\cite{Dinh_2023_latticematch_AlScN_MBE}. Moreover, the change in $a$, $c$ constants and $c/a$ ratio observed in this work also aligns with the values reported by Dinh et al for up to $x$ = 0.15\cite{Dinh_2023_latticematch_AlScN_MBE}, demonstrating the similarities in these MBE AlScN thin films. On the other hand, in recent works on MBE AlScN films, Motoki et al\cite{motoki_improved_2024} and Kumar et al\cite{kumar2024pinpointinglatticematchedconditionswurtzite} suggested the lattice matched Sc composition of 12\% by fitting the reported in-plane lattice constant $a$ in literature and 14\% by RSM analysis, respectively. The difference in growth temperature, ranging from 550 $^\circ$C \cite{kumar2024pinpointinglatticematchedconditionswurtzite} to 700 $^\circ$C\cite{Dinh_2023_latticematch_AlScN_MBE}, could impact the lattice constants,\cite{Dzuba_2022_Elimination_phaseimpurity_AlScN} and, therefore, the lattice-matched composition. Recent studies also reported Sc atomic segregation in both MBE\cite{Dzuba_2022_Elimination_phaseimpurity_AlScN, ndiaye_alloy_2023} and sputtered\cite{zhang_nanoscale_2023} AlScN. Atomic segregation can induce local variation of lattice constants to different extents, impacting the lattice matched composition. One variable of MBE growth not examined in this work is the effect of III/V ratio on structural properties, which could affect the lattice parameters and surface morphology of AlScN\cite{Hardy_2017_epitaxial_ScAlN}. Nevertheless, the wide variation in reported lattice-matched composition means that careful calibrations are needed to determine the correct lattice-matched Sc content depending on growth techniques and conditions. 

\begin{figure*}[t]
\includegraphics[width=\textwidth]{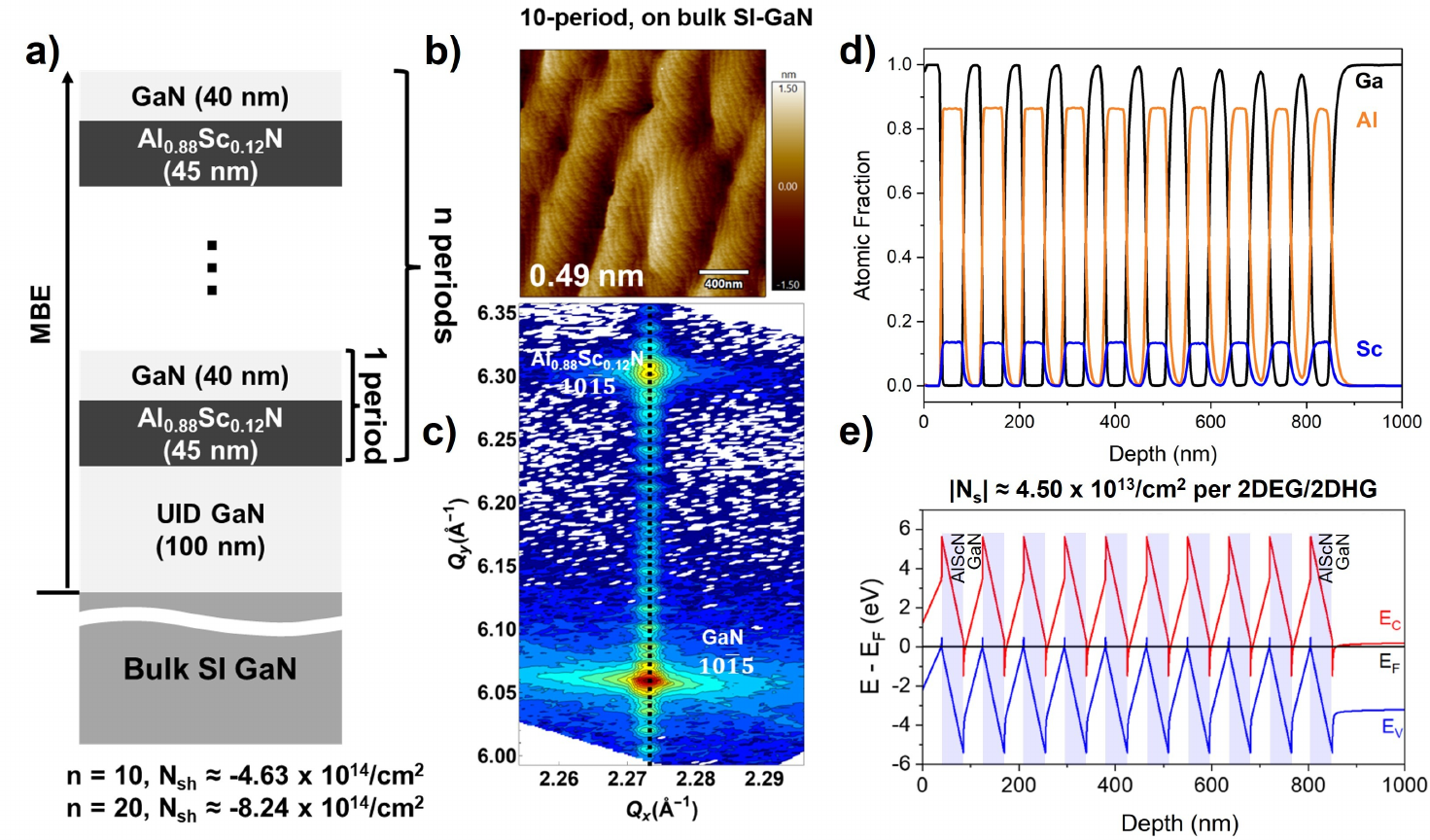}
\caption{(a) Heterostructure schematics of n-period (n = 10, 20) AlScN/GaN multilayer structures and their corresponding Hall effect measurement sheet carrier densities. (b) 2$\times$2 $ \mu$m$^2$ AFM micrograph of the 10-period sample shows step-flow growth mode was maintained even with nitrogen-rich AlScN growth and a total MBE thickness of nearly 1 $\mu$m; sub-5 $\Angstrom$ RMS roughness was achieved, showing the high layer and interface qualities enabled by near lattice-matched growth conditions. Similar surface morphology and roughness were seen on the 20-period sample (supplementary, Figure~S1). (c) The RSM scan of the 10-period sample shows that all MBE layers are coherently strained to the bulk GaN substrate; this is enabled by the higher critical AlScN thickness for near lattice-matched Sc compositions. (d) SIMS data of the 10-period sample, showing stable Sc composition (13 $\pm$ 1\%) was achieved in AlScN layers throughout the heterostructure compared to 12 $\pm$ 1.5\% measured by EDS.} (e) Schrodinger-Poisson 1D Simulation of the 10-period heterostructure, showing ten 2DEGs at the AlScN-GaN interfaces and ten 2DHGs at the GaN-AlScN interface with 4.5 $\times$ 10$^{13}$/cm$^2$ sheet charge density per 2DEG or 2DHG. AlScN layers are highlighted in blue stripes. From the experimental sheet charge densities and 1D Poisson simulation, we note that the net experimental sheet charge density scales almost linearly with the number of AlScN/GaN periods.  
\label{fig:Figure 5}
\end{figure*}


Figure~\ref{fig:Figure 4} summarizes the effect of growth temperature on the structural properties of single layer, 100-nm AlScN thin films grown on bulk n$^{+}$GaN substrate. Hardy et al and Dzuba et al both reported a strong dependence of lattice constants on growth temperature\cite{Hardy_2020_phasepurity_highSc_ScAlN_MBE, Dzuba_2022_Elimination_phaseimpurity_AlScN}. Here, we grew AlScN with three different Sc compositions (9\%, 13\%, 19\%) under nominally similar conditions and changed the substrate temperature between 330~$^{\circ}$C and 630~$^{\circ}$C. Figures~\ref{fig:Figure 4}(a) and \ref{fig:Figure 4}(b) show the change in $a$ and $c$ with respect to growth temperature, respectively. From Figure \ref{fig:Figure 4}(a), we see that the 9\% Sc AlScN films were almost pseudomorphically grown on GaN with no change in the $a$ parameter as a function of substrate temperature. On the other hand, the 19\% Sc containing AlScN samples are fully relaxed, exhibiting $\sim$ 1.5\% in-plane lattice mismatch compared to GaN. Notably, the $a$ lattice constant exhibited a strong temperature dependence in 13\% samples. For $T_{sub}$ = 330~$^{\circ}$C and 430~$^{\circ}$C, the 13\% Sc AlScN films were nearly fully strained to GaN, i.e. $a_{AlScN}$ $\sim$ 3.19 $\Angstrom$. However, $a$ quickly increased with increasing substrate temperature, approaching $a_{AlScN}$ $\sim$ 3.23 $\Angstrom$ for $T_{sub}$ = 630~$^{\circ}$C. This increase is induced by the elevated effects of strain relaxation at higher growth temperatures. Therefore, Figure~\ref{fig:Figure 4}(a) further corroborates the lattice-matched composition to be near 9-11\% Sc. Interestingly, the $a$ lattice parameter of the 9\% Sc AlScN film is nearly equal to the $a$ lattice parameter of GaN [Figure~\ref{fig:Figure 4}(a)], whereas the $c$ lattice parameter of the 9\% Sc AlScN film is nearly equal to the $c$ lattice parameter of AlN [Figure~\ref{fig:Figure 4}(b)], suggesting the ability to lattice match 9\% Sc AlScN to c-plane GaN or m-plane AlN.

It is interesting to note from Figure \ref{fig:Figure 4}(a) that while the $c$ parameter of the 9\% samples is not strongly dependent on the growth temperature, the $c$ parameter of the relaxed 19\% samples decreases sharply with increasing growth temperature, suggesting a decrease in unit cell volume with increasing growth temperature. The same decrease in $c$ parameter with increasing growth temperature was also reported by Wang et al for 50-100 nm MBE 20\% Sc AlScN\cite{Wang_2020_MBE_characterization_ScAlN} and by Dzuba et al for 40 nm MBE 14\% and 18\% Sc AlScN films.\cite{Dzuba_2022_Elimination_phaseimpurity_AlScN} Both works attributed this effect to residual strain induced by different growth temperatures, again suggesting that the lattice-matched composition is lower than 18\% Sc. Figure \ref{fig:Figure 4}(c) shows the change in surface morphology of AlScN thin films with Sc composition and growth temperature. For nearly pseudomorphic AlScN on GaN samples, the AFM surface morphology showed larger but smoother islands that could follow the underlying GaN atomic steps, and the rms roughnesses were below 0.35 nm. Higher Sc composition and growth temperature induced stronger strain relaxation and promoted the formation of smaller, more three-dimensional islands on the surface, resulting in rms roughnesses above 0.4 nm.

From the Sc composition- and temperature-dependent data described, the optimal growth conditions of lattice-matched AlScN on GaN for multilayer structures are 9-11\% Sc at growth temperatures between 530$^{\circ}$C and 630$^{\circ}$C. Using the optimized growth conditions of single layer AlScN/GaN, we grew near lattice-matched multilayer AlScN/GaN heterostructures with 45 nm AlScN/40 nm GaN periods and a targeted 12\% Sc composition for structural and transport properties characterization. By using semi-insulating GaN substrates, we were able to measure the total polarization induced mobile charges formed at the AlScN/GaN and GaN/AlScN heterointerfaces. Similar heterostructures grown on conductive n$^{+}$GaN substrates have been shown to be promising new lattice-matched GaN-based distributed Bragg reflectors\cite{van_deurzen_2023_AlScN_DBR}, but the transport properties of AlScN/GaN conductive channels were not reported due to n-type conductivity of the substrate. A ten period (n = 10) and a twenty period (n = 20) AlScN/GaN (45/40 nm) heterostructure were grown at $T_{sub}$ = 530~$^{\circ}$C, as shown in Figure \ref{fig:Figure 5}(a). While this growth temperature is 100-150~$^{\circ}$C below the optimal growth temperature of GaN in our MBE system, it is more suitable for near lattice-matched AlScN growths as depicted in Figure \ref{fig:Figure 4}. The AFM image in  Figure \ref{fig:Figure 5}(b) shows step-flow growth mode and clear atomic steps were achieved in both samples, with rms roughness of 0.49 nm for n = 10 and 0.54 nm for the n = 20 (supplementary, Figure~S1) despite the N-rich growth condition of the AlScN layers and the relatively low growth temperature of the GaN layers. 

\begin{figure*}[t]
\includegraphics[width=\textwidth]{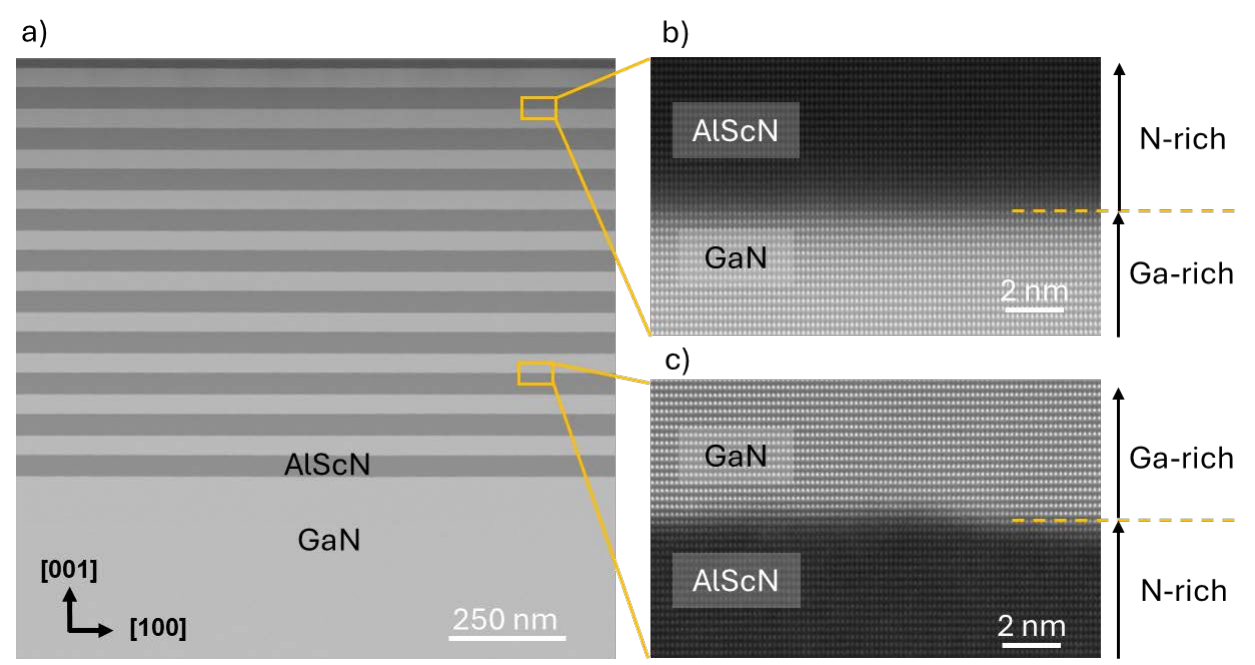}
\caption{ (a) Large field-of-view 
ADF-STEM}
image of 
12\% Sc AlScN/GaN ten period multilayer structure on bulk GaN substrate. (b) and (c) Atomic resolution ADF-STEM image of representative AlScN/GaN interfaces and associated MBE growth conditions.
\label{fig:Figure 6}
\end{figure*}


The RSM in Figure \ref{fig:Figure 5}(c) reveals that the AlScN layers are nearly strained to the underlying GaN substrate, showing a 0.02\% higher in-plane lattice constant compared to GaN. This mismatch is increased to 0.03\% for n = 20, suggesting a stronger compressive strain relaxation of AlScN layers, likely due to the Sc composition slightly exceeding the lattice-matched composition (supplementary, Figure~S1). The well-resolved interference thickness fringes in RSM scans confirm the high crystalline and interface quality of AlScN and GaN layers. Symmetric 2$\theta$-$\omega$ scans of both samples show strong Pendell\"{o}sung fringes, suggesting sharp interfaces between the AlScN and GaN layers (supplementary, Figure~S1). In situ RHEED tracking corroborated this, showing streaky RHEED patterns along the <110> azimuth of AlScN throughout the growth (supplementary, Figure~S2). The smooth surface morphology with atomic steps and high crystalline and interface quality is enabled by targeting the lattice-matched composition of 11\% Sc. XRD fringe simulation of both samples suggests the AlScN and GaN thickness to be 45 nm and 40 nm (supplementary, Figure~S1), respectively, agreeing well with the targeted thicknesses. 

A strong correlation between the XRD rocking curve (RC) full-width-at-half maximum (FWHM) broadening and strain relaxation due to lattice mismatch was found in previous MBE AlScN studies.\cite{Dinh_2023_latticematch_AlScN_MBE, kumar2024pinpointinglatticematchedconditionswurtzite} For n = 10, RC scans about the symmetric [0002] azimuth and asymmetric [10$\bar{1}$5] azimuth of the AlScN peak RC exhibited full-width-at-half maximum FWHM values of 30 arcsec and 30 arcsec, respectively. These low values closely follow the substrate RC FWHMs of 27 arcsec and 30 arcsec for the same azimuths (supplementary, Figure~S3). However, we measured an increase in the [10$\bar{1}$5] azimuth RC FWHM to 48 arcsec for n = 20, corresponding to the increased strain relaxation observed in RSM (supplementary, Figure~S1). ToF-SIMS data of the ten period sample in Figure \ref{fig:Figure 5}(d) demonstrates consistent Sc composition of 13 $\pm$ 1\% in AlScN layers throughout the heterostructure, agreeing with the 12 $\pm$ 1.5\% Sc composition measured in both multilayer structures by EDS. The atomic composition tailing between bottom GaN and AlScN layers is likely an artifact of SIMS spatial resolution. This composition is above the lattice-matched composition range, thereby inducing the increased strain relaxation observed from the ten-period to the twenty-period sample. While we clearly observed strain relaxation in 100-nm single layer AlScN films with Sc composition above 11\% (supplementary, Figure~S1), the critical thickness in a superlattice heterostructure can be four times higher than that of single layers due to the increased resistive force of misfit dislocation formation at new interfaces, resulting in thick pseudomorphic heterostructures with lattice-mismatched 12\% Sc AlScN-GaN epilayers.\cite{matthews_defects_1974}

Using 1D Poisson\cite{gregsnider1DP}, a self-consistent one-dimensional Schrodinger-Poisson solver, the energy band diagram of the ten period AlScN/GaN sample was simulated, as shown in Figure \ref{fig:Figure 5}(e). The 1D Poisson simulation predicted ten two-dimensional electron gases (2DEGs) and ten two-dimensional hole gases (2DHGs) to form at the AlScN/GaN and GaN/AlScN interfaces, respectively. The absolute carrier density for each 2DEG or 2DHG is |$N_{sh}$| = 4.5 $\times$ 10$^{13}$/cm$^2$. While 2DHGs have not been reported in AlScN/GaN heterostructures, their formation and contribution can not be experimentally ruled out.

Hall effect measurements at 300 K revealed extremely high mobile electron density in both samples [Figure~\ref{fig:Figure 5}(a)], with experimental net sheet carrier density $N_{sh}$ = - 4.63 $\times$ 10$^{14}$/cm$^2$ for n = 10 and $N_{sh}$ = - 8.24 $\times$ 10$^{14}$/cm$^2$ for n = 20, both of which are the highest values reported in multichannel GaN heterostructures, with less than 6\% of the measured net sheet charge density is expected to come from n-type conductivity of the unintentionally doped GaN layers. The negative Hall coefficients indicated that the majority of carriers are electrons. By dividing the measured carrier density by the simulated 2DEG density of 4.5 $\times$ 10$^{13}$/cm$^2$ per channel, we get 10.2 effective 2DEGs for n = 10 and 18.4 effective 2DEGs for n = 20, thereby demonstrating linear scaling of net charge density with the number of AlScN/GaN periods. The electron mobilities measured for n = 10 and n = 20 samples are 14 cm$^2$/V.s and 2 cm$^2$/V.s, respectively. Since excess Ga was fully desorbed before N-rich AlScN growths, strong interface roughness scattering and alloy scattering are expected in these samples. Interface roughness scattering and alloy scattering have a strong effect on electron mobility at the AlScN-GaN interface\cite{casamento_transport_2022}, so growth optimization is needed to enhance the sheet charge mobility. Specifically, the introduction of an AlN interlayer at each AlScN-GaN interface can help boost the 2DEG mobility\cite{casamento_transport_2022}. However, AlN and AlScN strain engineering will be needed to maintain the pseudomorphic strain throughout the heterostructure with multiple AlN interlayers. Furthermore, the possible contribution of heavier carriers and lower mobility from 2DHGs could also limit the Hall effect mobility measured in these samples, suggesting the potential need for Si delta doping at the GaN-AlScN interfaces.

Figure \ref{fig:Figure 6}(a) displays an annular dark-field scanning transmission electron microscopy (ADF-STEM) image showing ten uniform layers of the heterostructure composed of 45 nm thick AlScN with 12\% Sc and 40 nm thick GaN.  No threading or edge dislocations are visible in this field of view ($\sim$1.3 $\mu$m) indicating high quality epitaxial growth near lattice matched conditions between AlScN and GaN. In a larger field-of-view ADF image (spanning more than 7$\mu$m), we observed vertical dislocations stemming from the substrate and buffer layer interface. These dislocations  contribute to the relaxation of the slight remnant strain present between AlScN and GaN (supplementary, Figure S4). The upper bound of in-plane strain calculated based on the average distance between visible dislocations corresponds to 0.02\% compressive strain of AlScN epilayers, consistent with RSM measurements. The interfaces between GaN and AlScN layers were studied by atomic resolution ADF-STEM images. A smooth diffuse AlScN/GaN interface is observed across 3-4 monolayers in Figure \ref{fig:Figure 6}(b). The gradient in the image contrast describes the change in chemical composition despite our efforts during growth to desorb excess Ga and minimize the Ga incorporation into AlScN. The physical origin of the diffuse interface requires further investigation to minimize the alloy scattering expected to affect the 2DEG. In comparison, Figure \ref{fig:Figure 6}(c) shows $\sim$1nm surface roughness at the GaN/AlScN interface. This interface morphology and roughness are consistent with the surface morphology observed in other AlScN epilayers grown under nitrogen-rich conditions.\cite{Hardy_2020_phasepurity_highSc_ScAlN_MBE}

We further analyzed the surface layers of the ten period 12\% Sc composition AlScN/GaN multilayer sample using phase lock-in analysis \cite{goodge_phase_lockin_2022}. We did not observe in-plane lattice mismatch within the precision of the measurement across an $\sim$80 nm field of view (supplementary, Figure S5). The strain variance in AlScN is on the same order as the noise variance in GaN and does not display systematic differences. The AlScN out-of-plane lattice constant matches the $c$ lattice constant of 12\% Sc composition AlScN film [Fig. \ref{fig:Figure 3}(b)]. We therefore conclude that the AlScN layers are strained to the same in-plane lattice constant as GaN in the ten-period multilayer heterostructure despite the slight variation from lattice-matched Sc composition.

As a final point of discussion, lattice-matched AlScN/GaN multilayer structures have a bright outlook for future GaN-based electronics, optoelectronics, and photonics. The pseudomorphic nature of lattice-matched epilayers also means that the number of AlScN/GaN periods and AlScN thickness can be flexibly chosen for various multilayer heterostructure engineering. For example, lattice-matched AlScN/GaN distributed Bragg reflectors with AlScN layers thicker than 40 nm can be a new platform for GaN-based bottom DBRs that can outperform the existing AlInN/GaN DBRs\cite{van_deurzen_2023_AlScN_DBR}. AlScN/GaN multiple quantum wells with AlScN layers thinner than 10 nm are capable of strong near-infrared intersubband absorption (ISBA)\cite{Dzuba_2022_Elimination_phaseimpurity_AlScN, gopakumar_conduction-band_2024}. Similarly, GaN-based heterostructures that require Al(Ga,In)N/GaN multiple quantum wells, such as resonant tunneling diodes, quantum cascade lasers, and multichannel transistors, stand to benefit from the wide bandgap, highly polar lattice-matched AlScN material. The understanding of lattice parameters as a function of alloy composition and growth conditions also provides more general guidance for strain engineering of AlScN in multilayer\cite{zheng_electrical_2021}, high Sc composition\cite{hardy_nucleation_2023}, or compositionally graded\cite{simon_polarization-induced_2010} heterostructures. It is important to note that while using the same growth technique (MBE), the lattice-matched composition of AlScN in this work, Dinh et al's, Kumar et al's, and Dzuba et al's are 9--11\%, 9\%, 14\%, and 18\% Sc, respectively.\cite{Dinh_2023_latticematch_AlScN_MBE,kumar2024pinpointinglatticematchedconditionswurtzite,Dzuba_2022_Elimination_phaseimpurity_AlScN} Therefore, a more comprehensive examination on the effects of growth conditions on the AlScN lattice-matched composition is needed to deliver high quality pseudomorphic AlScN/GaN multilayer heterostructures.

\section{\label{sec:level4}Conclusion}

In summary, new understandings of AlScN/GaN single layer and multilayer heterostructures were achieved in this work. First, we found that the lattice-matched composition of MBE-grown AlScN on GaN is around 9-11 \% Sc. Lattice-matched AlScN films with 9--11\% Sc can be grown using a large temperature window, providing guidance to optimizing the growth conditions and crystal quality of AlScN epilayers and AlScN/GaN multilayer structures. The lower lattice-matched composition here can have important implications for AlScN integration with GaN, such as interplaying strain engineering, high-k dielectric and ferroelectric properties. Next, near pseudomophic AlScN/GaN multilayer heterostructures with ten and twenty AlScN/GaN periods were realized by MBE. Excellent surface and interface roughness were verified via AFM, XRD, and ADF-STEM analyses. Net mobile electron charge density as high as 8.24 $\times$ 10$^{14}$/cm$^2$ for twenty channels and linear carrier density scaling with the number of AlScN/GaN periods was achieved, suggesting the presence of multiple polarization-induced conduction channels enabled by growing near lattice-matched conditions. These results provide a new understanding of AlScN/GaN epitaxial growth by MBE and highlight the benefits of lattice-matched AlScN and lattice-matched AlScN/GaN multilayer heterostructures for multiple quantum well heterostructures in future GaN-based power electronics and photonics.

\section*{Supplementary Material}
See the supplementary material for XRD, AFM, RSM, and XRD rocking curve scans of the ten and twenty period AlScN/GaN samples on bulk semi-insulating GaN substrates, the RHEED patterns of the AlScN during a multilayer growth, the RSM of single layer AlScN films with various Sc compositions, the ADF-STEM image with vertical dislocation stemming from the substrate, and the RBS spectra.

\section*{Acknowledgements}
The authors acknowledge the use of facilities and instrumentation supported by NSF through the Cornell University Materials Research Science and Engineering Center DMR-1719875 and the CESI Shared Facilities partly sponsored by NSF No. MRI DMR-1631282. The Thermo Fisher Spectra 300 X-CFEG was acquired with support from PARADIM, an NSF MIP (DMR-2039380) and Cornell University. This work was supported in part by SUPREME, one of seven centers in JUMP 2.0, a Semiconductor Research Corporation (SRC) program sponsored by DARPA. This material is based upon work supported by the Air Force Office of Scientific Research and the Office of Naval Research under award number FA9550-23-1-0338. Any opinions, findings, and conclusions or recommendations expressed in this material are those of the author(s) and do not necessarily reflect the views of the United States Air Force or the Office of Naval Research. JJPC and RSG gratefully acknowledge support from the National Science Foundation (Grant No. DMR-1810280) and ECCS-BSF-2240388. We also acknowledge staff support from the Michigan Ion Beam Laboratory at the University of Michigan.

\section*{Author Declarations}
\subsection*{Conflict of Interest}
The authors have no conflicts to disclose.
\section*{Data Availability}

The data that support the findings of this study are available from the corresponding author upon reasonable request.


\bibliography{AlScNmultilayergrowth_ref}

\end{document}


\graphicspath{{./figures/}}

\title{\large{Supplementary Information - Lattice-Matched Multiple Channel AlScN/GaN Heterostructures}}

\author{Thai-Son~Nguyen}
\email[Electronic mail: ]{tn354@cornell.edu}
\affiliation{Department of Materials Science and Engineering, Cornell University, Ithaca, New York 14853, USA}
\author{Naomi~Pieczulewski}
\affiliation{Department of Materials Science and Engineering, Cornell University, Ithaca, New York 14853, USA}
\author{Chandrashekhar~Savant}
\affiliation{Department of Materials Science and Engineering, Cornell University, Ithaca, New York 14853, USA}
\author{Joshua~J. P. Cooper}
\affiliation{Department of Materials Science and Engineering, University of Michigan, Ann Arbor, Michigan 48109, USA}
\author{Joseph~Casamento}
\affiliation{Department of Materials Science and Engineering, Massachusetts Institute of Technology, \\Cambridge, Massachusetts 02139, USA}
\author{Rachel~S. 
Goldman}
\affiliation{Department of Materials Science and Engineering, University of Michigan, Ann Arbor, Michigan 48109, USA}
\author{David A.~Muller}
\affiliation{Kavli Institute at Cornell for Nanoscale Science, Cornell University, Ithaca, New York 14853, USA}
\affiliation{School of Applied and Engineering Physics, Cornell University, Ithaca, New York 14853, USA}
\author{Huili G.~Xing}
\affiliation{Department of Materials Science and Engineering, Cornell University, Ithaca, New York 14853, USA}
\affiliation{Kavli Institute at Cornell for Nanoscale Science, Cornell University, Ithaca, New York 14853, USA}
\affiliation{Department of Electrical and Computer Engineering, Cornell University, Ithaca, New York 14853, USA}
\author{Debdeep~Jena}
\affiliation{Department of Materials Science and Engineering, Cornell University, Ithaca, New York 14853, USA}
\affiliation{Kavli Institute at Cornell for Nanoscale Science, Cornell University, Ithaca, New York 14853, USA}
\affiliation{School of Applied and Engineering Physics, Cornell University, Ithaca, New York 14853, USA}
\affiliation{Department of Electrical and Computer Engineering, Cornell University, Ithaca, New York 14853, USA}

\begin{abstract}
The supplementary material includes:\\
Figure~S1 -- RSM of single layer AlScN with various Sc compositions, XRD, AFM and RSM of the ten and twenty period samples
Figure~S2 -- RHEED pattern of the AlScN layers during growth of the ten period sample
Figure~S3 -- XRD rocking curve (RC) scans comparing the AlScN (0002) and (10$\bar{1}$5) peaks to the GaN (0002) and (10$\bar{1}$5) peaks, respectively.
Figure~S4 -- ADF-STEM image showing the dislocations present in ten period 12\% Sc AlScN/GaN film.
Figure~S5 -- ADF-STEM image and lattice parameter analysis in AlScN/GaN multilayer heterostructure.
Figure~S6 -- RBS yield vs. backscattered energy spectra of AlScN/GaN films.
\end{abstract}

\maketitle


\section*{Supplementary Information}

\begin{figure}[H]\includegraphics[width=\textwidth]{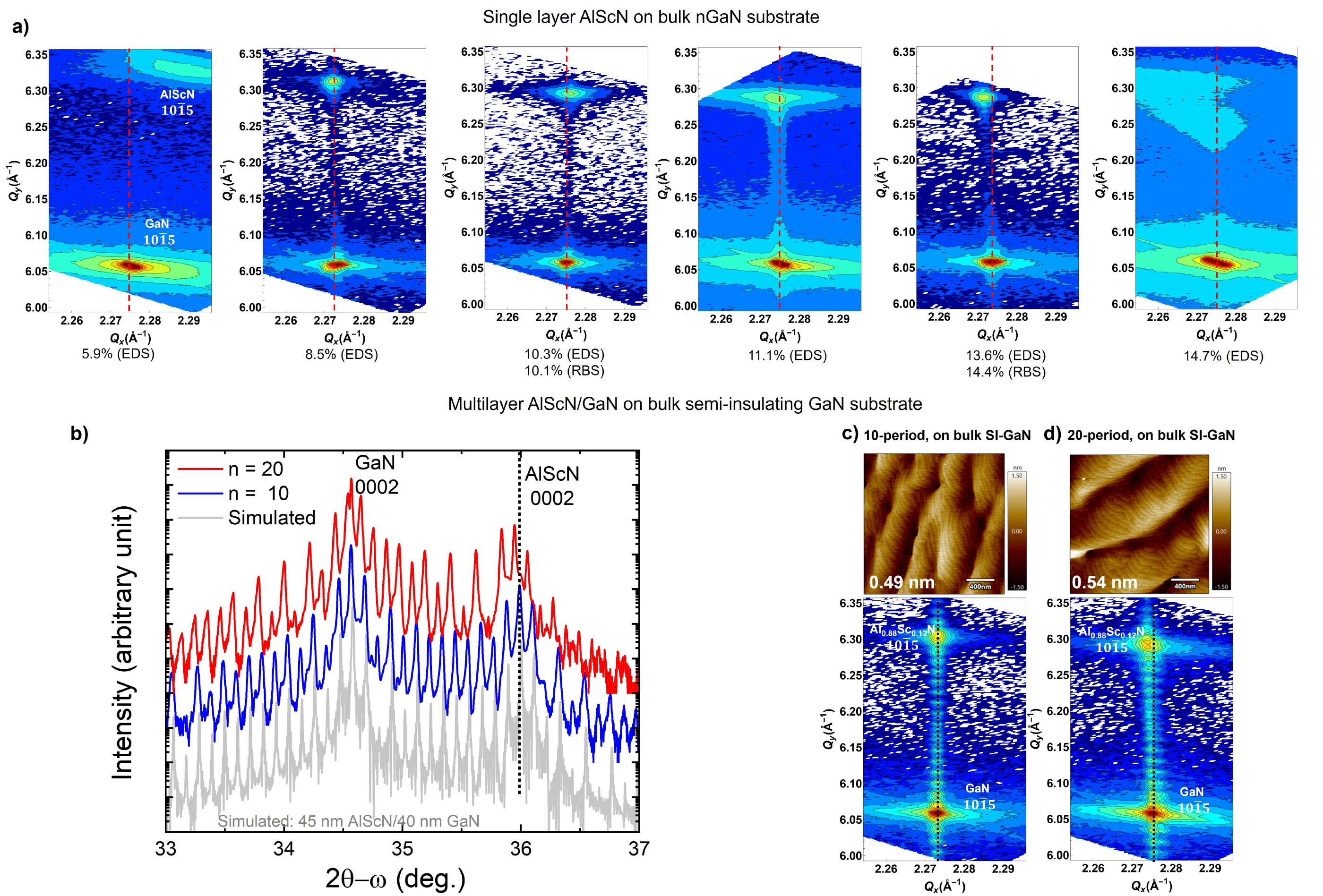}
\caption{(a) RSM scan along the (10$\bar{1}$5) azimuth of single layer AlScN on GaN with different Sc compositions, showing strain relaxation outside of the 8.5-11\% Sc composition as indicated by the AlScN peak shift on Q$_x$. (b) Symmetric 2$\theta$-$\omega$ scans show strong interference fringes and indicate the high interface quality between AlScN and GaN layers. A simulated structure of ten period 12\% Sc AlScN/GaN with AlScN and GaN thickness of 45 nm and 40 nm, respectively, accurately describes the interference fringe patterns, thereby indirectly confirming the intended layer thicknesses were achieved in these samples. (c) Post-growth 2$\times$2 $\mu$m$^2$ AFM micrograph of the ten period sample, showing clear atomic steps and rms roughness of 0.49 nm; RSM scan shows that the AlScN and GaN layers are nearly pseudomorphically strained to the GaN substrate with a 0.02\% in-plane lattice mismatch. (d) Post-growth $\mu$m$^2$ AFM micrograph of the twenty period sample, showing similar atomic steps and rms roughness of 0.54 nm; RSM scan suggests a 0.03\%  in-plane lattice mismatch between AlScN and GaN. The  While the surface morphology is smooth with atomic steps, wavy patterns were observed, likely due to the effect of step meandering of GaN grown at 530$^{\circ}$C.} 
\label{Figure_S1}
\end{figure}


\begin{figure}[H]\includegraphics[width=\textwidth]{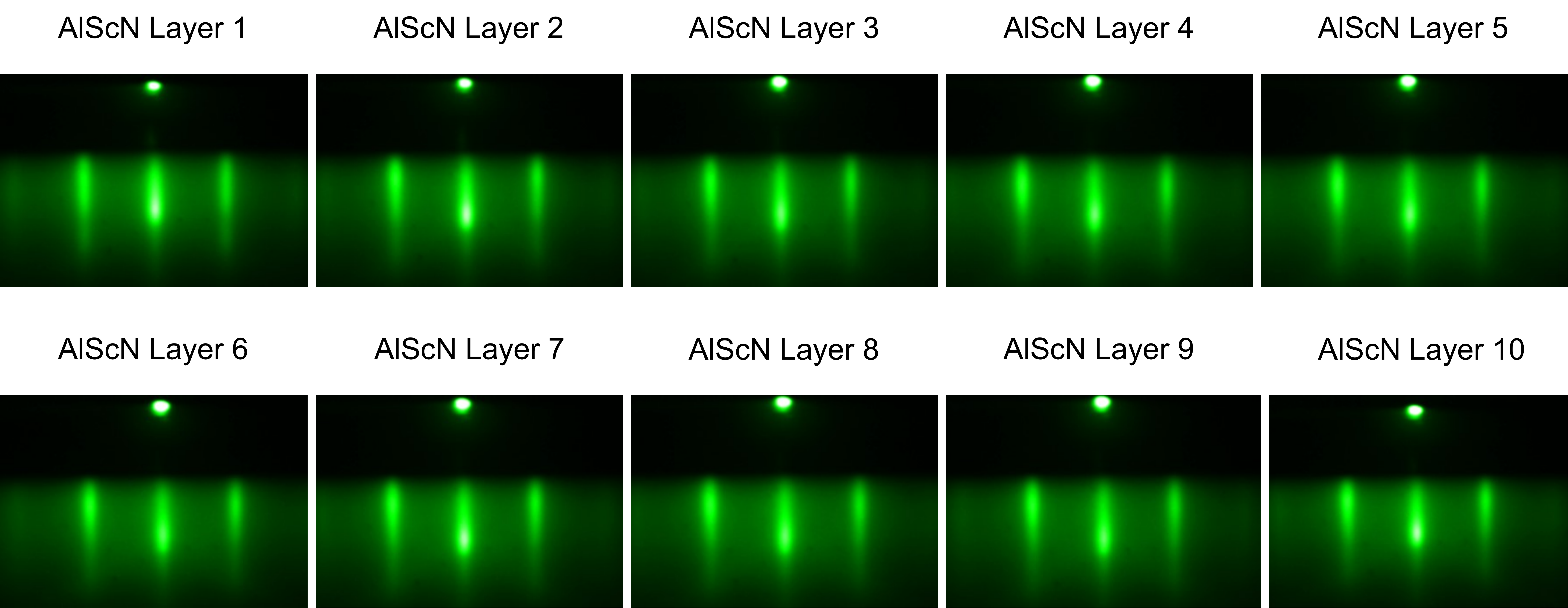}
\caption{In situ RHEED tracking along the <110> azimuth during growth of the ten period 45 nm/40 nm AlScN/GaN heterostructure at the end of each AlScN layer. In spite of the N-rich growth condition, the RHEED pattern remained streaky throughout the growth, corroborating the smooth surface feature with atomic steps measured by XRD and the strong interference fringes measured by symmetric 2$\theta$-$\omega$ and RSM scans.} 
\label{Figure_S2}
\end{figure}


\begin{figure}[H]\includegraphics[width=\textwidth]{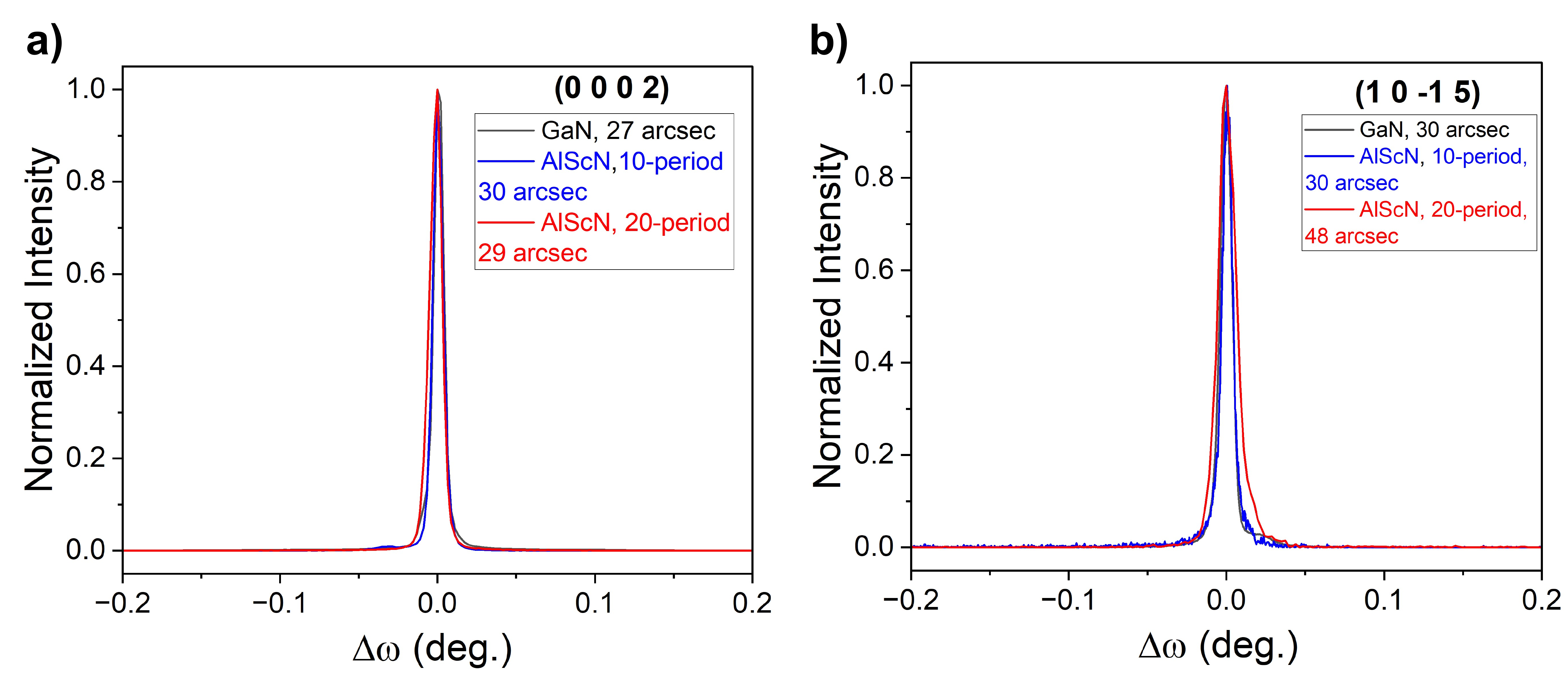}
\caption{XRD Rocking Curve (RC) scans of the AlScN (a) (0002) and (b) (10$\bar{1}$5) azimuths compared to RC of similar azimuths of the GaN substrate for the ten period and twenty period AlScN/GaN samples on GaN template. Thanks to the near lattice-matched Sc composition, high crystal quality along the c-axis was obtained in both samples, with RC FWHM around the (0002) azimuth of AlScN to be 30 (29) arcsec for n = 10 (20), closely matching that of the GaN substrate (27 arcsec). As shown in Supplementary Figure S1 and discussed in the main text, strain relaxation was observed for n = 20 due to deviation of Sc content from the lattice-matched composition. As a result, the RC along the asymmetric (10$\bar{1}$5) peak broadened significantly with FWHM = 48 arcsec. For n = 10, the structural quality remained high, with RC FWHM = 30 arcsec that matched the substrate’s FWHM. These RC data depict the importance of reaching lattice-matched conditions for growing thicker AlScN/GaN multilayer structures.} 
\label{Figure_S3}
\end{figure}


\begin{figure}[H]\includegraphics[width=\textwidth]{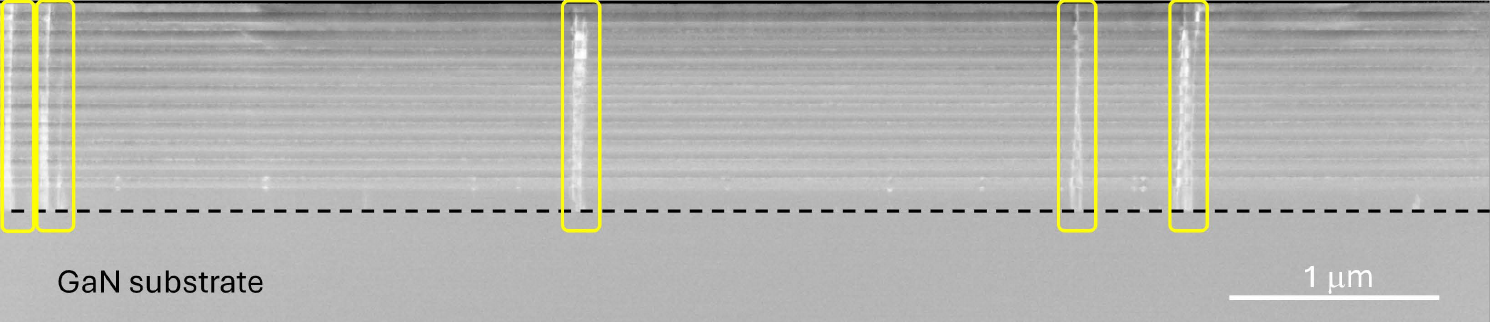}
\caption{Low-angle annular dark field (LAADF) STEM image of ten period 12\% Sc AlScN/GaN heterostructure on semi-insulating GaN substrate highlights dislocations present in the sample. The vertical dislocations boxed in yellow begin at the GaN substrate to GaN buffer interface indicated by the dashed line and run to the surface of the film. These dislocations are most likely due to insufficient surface cleaning of the bulk substrate before GaN buffer and epilayer growth. Although not caused by strain between AlScN and GaN, the dislocations present can relax the AlScN. Therefore, the average dislocation density ($\approx$ 1.4$\times$10$^{9}$ /cm$^2$) was used to calculate the upper bound of strain. The average in-plane strain in this film could be described as <0.02\%.} 
\label{Figure_S4}
\end{figure}
\begin{figure}[H]\includegraphics[width=\textwidth]{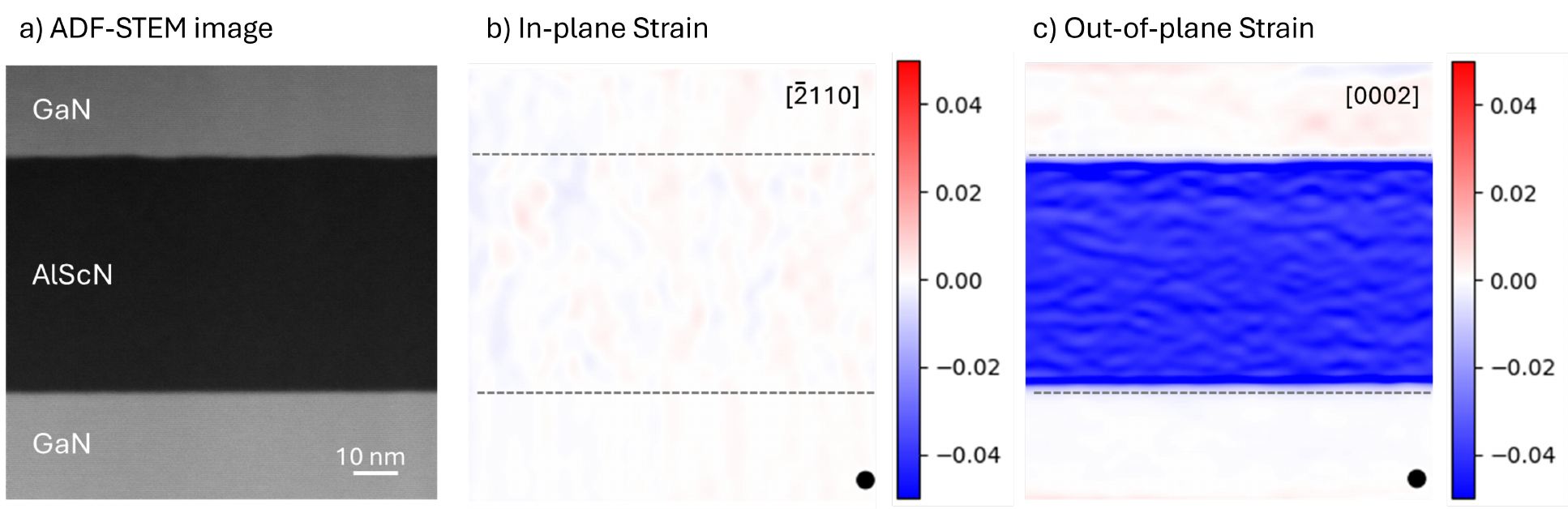}
\caption{a) ADF-STEM image of AlScN/GaN interfaces near the surface of 12\% Sc AlScN/GaN multilayer heterostructure. b), c) In-plane and out-of-plane strain calculated using phase lock-in analysis of ADF image in a). The strain is evaluated by applying a Gaussian mask around the specified reciprocal lattice vector resulting in real-space coarsening. The spatial resolution of the strain map dictated by the Gaussian mask is denoted by the black circle on the bottom right. The in-plane strain variation primarily displays the intrinsic noise consistent throughout both AlScN (standard deviation $\sigma$ = 0.0044) and GaN ($\sigma$ = 0.0032) layers. The out-of-plane strain in AlScN highlights the expected reduction in c axis lattice parameter. } 
\label{Figure_S5}
\end{figure}
\begin{figure}[H]\includegraphics[width=\textwidth]{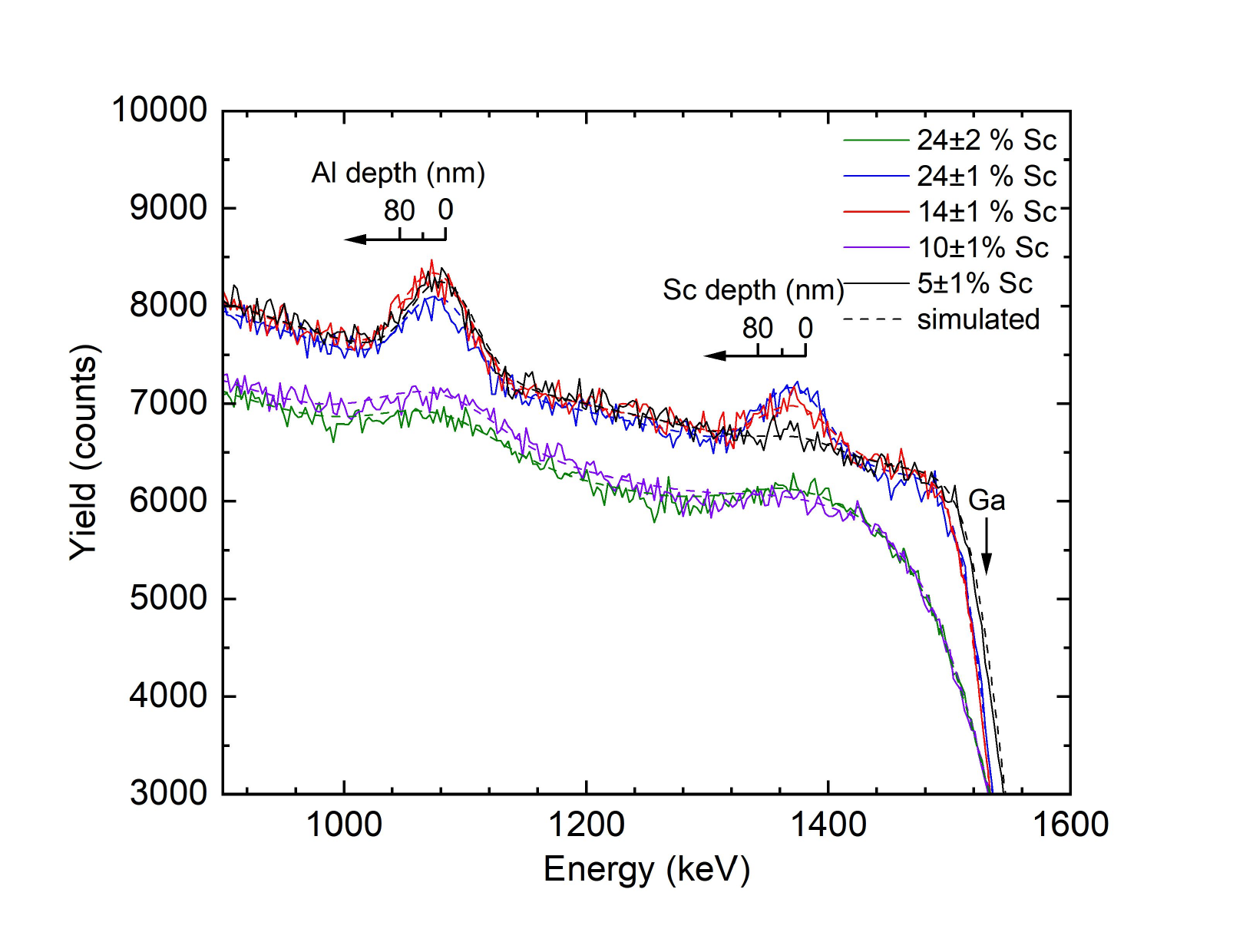}
\caption{(a) Rutherford backscattering spectrometry (RBS) yield vs. backscattered energy and depth for AlScN layers. RBS data are overlaid with SIMNRA fitted spectra assuming a uniform Sc depth profile, shown in solid lines and dashed lines, respectively. Although the spectra for the 24\% Sc (green) and 10\% Sc (purple) layers were collected with lower energy resolution (due to increased noise in the signal processing electronics), the accuracy of the compositional analysis is maintained.} 
\label{Figure_S6}
\end{figure}